\documentclass[final,3p,times,onecolumn]{elsarticle}

\usepackage{amssymb}
\usepackage{amsmath}
\usepackage{graphicx,stmaryrd}

\usepackage[utf8]{inputenc}
\usepackage[T1]{fontenc}
\DeclareUnicodeCharacter{2212}{\ensuremath{-}} 

\usepackage{tablefootnote} 

\usepackage{chemfig}
\usepackage[version=4]{mhchem}

\IfFileExists{newtxtext.sty}
   {\usepackage{newtxtext,newtxmath}}
   {\IfFileExists{stix.sty}
      {\usepackage{stix}}
      {\IfFileExists{mathptmx.sty}
      {\usepackage{mathptmx}}{} } }

\usepackage{textcomp}
\usepackage{hyperref}
\usepackage{tabularx}

\usepackage{xcolor}
\definecolor{LinkColor}{rgb}{0.256,0.439,0.588}

\hypersetup{
   colorlinks=true,
   linkbordercolor = {white},
   citecolor=blue,
   linkcolor={black},
   urlcolor=blue
   }

\journal{Journal of Alloys and Compounds}
\begin{document}

\begin{frontmatter}



\title{Fingerprint of $T_c$ advancement in Li-doped Bi-2223  superconductors prepared by  cationic  molecular mixing  within Pechini  sol-gel synthesis} 



\author[1]{ N.\ K.\ Man} 
\affiliation[1]{organization={School of Materials Science and Engineering, Hanoi University of Science and Technology}, \\  
            city={Hanoi},
            postcode={100000}, 
            country={Vietnam}}
\author[2,3]{Huu T.\ Do \corref{cor1}}
\ead{huutdo09@gmail.com}
\cortext[cor1]{Corresponding author}
\affiliation[2]{organization={Institute of Theoretical and Applied Research, Duy Tan University},
            city={Hanoi},
            postcode={100000}, 
            country={Vietnam}}

\affiliation[3]{organization={Faculty of Natural Sciences, Duy Tan University},
            city={Da Nang},
            postcode={550000}, 
            country={Vietnam}} 
\author[1]{ Nguyen V.\ Tu} 
\author[1]{Nguyen V.\ Quy} 

\begin{abstract}
Trilayered Bi-2223 superconductor  features the highest critical temperature $T_c$ among the  bismuth-based cuprate collection  
and     
symbolizes an ideal prototype for studying intrinsic superconducting properties.  
The previous solid-state reaction method substantiated the growth of the high-quality  Bi-2223 compounds  
but was accompanied by excessively laborious time and effort in terms of    multiple  grinding, pressing,  
as well as  calcining stages, 
so  finding a  less tedious synthesis path  is imperative. 
Here, we present an advanced sol-gel synthesis for assembling
the multicomponent complexity of  \ce{Bi_{1.4}Pb_{0.6}Sr_2Ca_2(Cu_{1−x}Li_x)_3O_{10+\delta}} superconductors  (Li-doped Bi-2223),  with $x$ = 0.0--0.20,   utilizing  metallic cationic  molecular mixing   
within the chemical Pechini polyesterization  route followed by single-step pyrolysis and sintering  stages. 
Although monovalent cations such as Li$^+$ pose limitations in establishing a perplex crosslinking network or chelating mechanism in the Pechini method, 
they represent a unique probe to elucidate the major chemical process during   polymerization. 
We observe that a 5 molar~\% Li-doped  sample pronounces the highest $T_c$ = 111.4 K among the series of  samples, as confirmed by both ac susceptibility and dc resistivity measurements, and is equivalent to the value obtained by our preceding solid state fabrication. 
In addition, we showcase a rare observation of layer-by-layer crystalline phase growth  under  microstructure probe. 
Through analyzing the reliable ac susceptibility data at low magnetic fields  in a wide range of frequency, we  provide  the quantum flux formation and flux creep mechanism by Anderson-M\"uller's model and Cole-Cole plot. 
 Our study paves a  perspicuous pathway for finding effective  fabrication for the high-quality Bi-2223 compounds as well as impurity-doped counterparts, by utilizing the appropriate synthesizing and doping processes. 

\end{abstract}

\begin{graphicalabstract}
\includegraphics[scale=0.3]{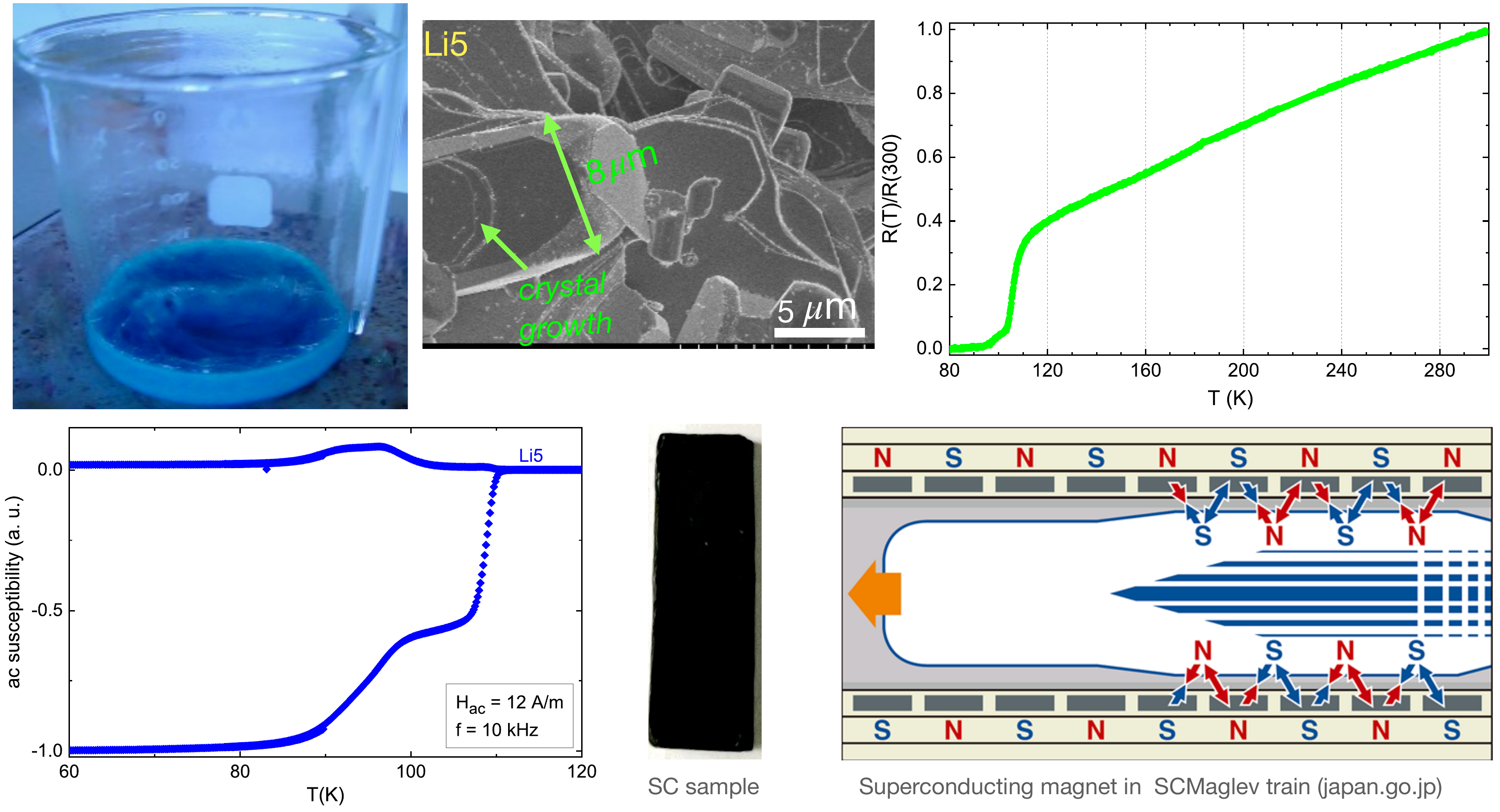}
\end{graphicalabstract}



\end{frontmatter}



\section{Introduction}
\label{sec1}

The dawn of high-$T_c$ superconductivity ignited in partial Ba-replaced La-Cu-O ceramic compounds containing mainstream copper-oxide ingredients, 
CuO$_2$  layers, which were categorized as cuprate materials   
 and pronounced the first observation of the transition temperature $T_c$ = 35 K \cite{Nobel-HTCS} above conventional Nb-based metallic alloys \cite{PhysRevLett.6.89, AKUNE20002095, PhysRevB.84.224518, Valente-Feliciano_2016}.  
 Following the class of rare earth RE-Ba-Cu-O materials emerges as $T_c$ = 90 K \cite{YBCO-1987}, featuring the first superconducting material that works above the abundant liquid nitrogen of 77 K, 
 but accompanied by a typically weak chemical and structural stability due to oxygen loss, losing superconductivity, moisture corrosion, and expensive RE-containing \cite{ZHOU1997223, ROA20124256}.
 The most stable group of  cuprate compounds without having RE or toxic components refers to the bismuth-based family  generalized by the chemical formula Bi$_2$Sr$_2$Ca$_{n-1}$Cu$_n$O$_{2n+4}$ (BSCCO) 
  symbolizes an ideal platform for chemical and structural stability escorted with $T_c$ pursuing a bell-curve shape with respect to the number of cuprate layers, such as 36 K for  single-layered Bi-2201 (n = 1), elevating to double-layer of 92 K for Bi-2212 (n = 2), reaching 110 K for Bi-2223 (n = 3), but drastically downturn to 90 K for Bi-2234 (n = 4) \cite{Maeda_1988, PhysRevB.38.2504, Science-2023-Bi2223, Rainer}.
However, the BSCCO family, especially for the triple-layered Bi-2223 materials, faces  a multitude of difficulties in fabricating high-quality samples in  single-phase or homogeneous single crystal forms  
while they often consist of  unwanted impurities pervading in the grain boundary \cite{RMP-2002-Hilgenkamp, Shen_2024}.    
Therefore, considerable experimental efforts have been employed to elucidate perspicuous pathways to synthesize high-quality Bi-2223 compounds   
as well as to raise their critical temperatures higher, typically  renovating a conventional solid-state \cite{PRM-2024-Do, RSI-2025-Do, CAO202412212}, 
performing atomic mixing coprecipitation  \cite{nano13152197, ZHANG2021103245}, spay-pyrolysis \cite{TSUDO19915, CAO202510014},   wire or tapes \cite{Bi-2223-wire, OGIWARA19971},  growing single crystal form  \cite{PhysRevB-O2-hole, Clayton}, 
chemical sol-gel \cite{RUBESOVA2012448, Shamsodini_2019, FALLAHARANI2022163201} and subsequently fabricating thin film \cite{ Apply-Phys-Let-2022, Shen_2024}.
Due to the intrinsic complexity of the multicomponent BSCCO category, 
the conventional solid-state reaction still retains the most widespread auspicious way to grow a high-quality bulk sample in a  mass quantities for traditional  known applications such as superconducting bulk magnets or wires  \cite{CHANG2025100182}, 
but it is intractable to  fabricate fine single crystal, thin film, and other superconducting devices for various appliances, 
such as high-$T_c$ superconducting terahertz devices and quantum interference instruments  \cite{Clayton, Shamsodini_2019, PhysRevApplied.22.044022, Tsujimoto_2025, PhysRevApplied.11.044004}. 

Among various atomic mixing techniques, sol-gel engages a promising approach; 
however, a traditional sol-gel synthesis relying on metallic alkoxide precursors has not been proven    to be  relevant and  useful for preparing a complicated multicomponent BSCCO sample caused by a substantial   hydrolysis rate difference of Bi, Pb, Sr, Ca and Cu alkoxides as well as the poor solubility of Cu-alkoxide \cite{RUBESOVA2012448, sol-gel-found-1996}. 
A state-of-the-art sol-gel technique based on a polymerization or polyesterization process, 
renowned Pechini method, aims to provide better quality  through a selection of  chelating agents such as citric acid, acetic acid,  ethylenediaminetetraacetic acid, etc.\ which entrap metallic ions in a complex solution up to homogeneous gel formation  \cite{RUBESOVA2012448, sol-gel-found-1996, Sol-gel-1990} or starting from  different precursors  either nitrate, acetate salts, and carbonate \cite{Shamsodini_2019}.   
As in our search of several attempts on the  Pechini route,  only a few of them provided high-quality samples that exhibited a sharp  superconducting transition around  110 K such as growing single crystals from sol-gel precursor \cite{Clayton}, adding   TiO$_2$ nanoparticles into Bi-2223 ceramics  \cite{FALLAHARANI2022163201}, utilizing different salt precursors \cite{Shamsodini_2019} 
or the recent  fabrication of thin film from sol-gel synthesis \cite{Shen_2024};
however, detailed chemical formations have not been  intensively advocated.  
 In addition, $T_\textrm{c}$ = 110.5--111 K typifies the highest value recorded for pristine Bi-2223 material by  magnetic susceptibility measurements \cite{PhysRevB-O2-hole, Clayton, B-Liang-2004, PRM-2024-Do},  
 but the samples synthesized by the latest sol-gel technology have not surpassed to the optimum \cite{Shen_2024, RUBESOVA2012448, FALLAHARANI201921878}.

%


Inspiring the genesis of high-$T_c$ superconductivity in   cuprate compounds results in appending charge carriers either electrons or holes into Mott insulators, e.g.,  partially replacing external Ba$^{2+}$ for  La$^{3+}$ cations \cite{Nobel-HTCS} 
or tuning internal oxygen content  within the prototypical  La$_2$CuO$_4$ insulators \cite{Neutron-LaCuO4, Smith-nature-1991}. 
In accordance with the indispensable correlation between the fabrication--structure--property principle, varying  elemental components does not only modify   fabricating technology of Bi-2223 materials, 
but also alters the superconducting  properties. 
For instance, partial replacement of Pb for Bi reduces the sintering time and temperature as well as creating a favorable liquid phase Ca$_2$PbO$_4$ that  promotes high-$T_c$ Bi-2223 crystalline growth \cite{Hole-doped-PRB, Pb-doped-Bi2223, Doped-Pb1, PRM-2024-Do}. 
Intercalating silver oxide into the Bi-2223 frame decreases the sintering time as a catalytic agent, but improves high-$T_c$ grain connectivity \cite{ManP-Ag, PhysRevB.50.10218, Muller-1999}. 
Furthermore, partially replacing  transition metal oxides of Ni$^{2+}$, Fe$^{2+}$/Fe$^{3+}$, and Co$^{2+}$ for Cu$^{2+}$ cation in Bi-2223 compounds reduces $T_\text{c}$   but an appropriate content  enhances pining force and possible critical current density  \citep{Pop, Pop1, Co-doped-Pop, DOGRUER2022163445, AFTABI2022164455, SUHAIMI2025178448, HARABOR20192742}. 
It has been suggested that presenting  TiO$_2$ nanoparticles between grains improves the connectivity and alignments in the synthesized compounds but reduces the critical temperature \cite{FALLAHARANI2022163201}.
Doping Li$_2$O into Bi-2223 cuprate reduced the sintering temperature and showcased the sign of $T_c$ elevation observed in previous experiments \cite{PRM-2024-Do, Matsubara, MatsubaraI}.  
However, performance of Li-doped cases by atomic mixing sol-gel synthesis has not been processed. 
To the best of our knowledge, a comprehensive and thorough interpretation  of the chemistry involved in the complex organic polymeric  Pechini pathway to synthesize Bi-2223 sample is not apparently stated. 

In short, we report our detailed study of chemical evolution with direct substitution of  Li$^+$ for Cu$^{2+}$ cations within the notable Pechini sol-gel  route.  
After that, we demonstrate the characterizing technique to investigate the crystal structure,
morphological topography, as well as focused critical temperature and intrinsic superconducting properties in Li-doped Bi(Pb)-2223 frameworks. 
In further elucidating the flux creep properties, we utilized the Anderson-M\"uller's analysis and Cole-Cole plot applying to a wide range of frequencies to elucidate the flux creep formation.


\section{Enterprising synthesis method, structure--morphology--superconducting--property relationship and theoretical model of ac magnetic susceptibility}
\label{subsec1}

\begin{figure*}[ht]
\centering
\includegraphics[scale=0.25]{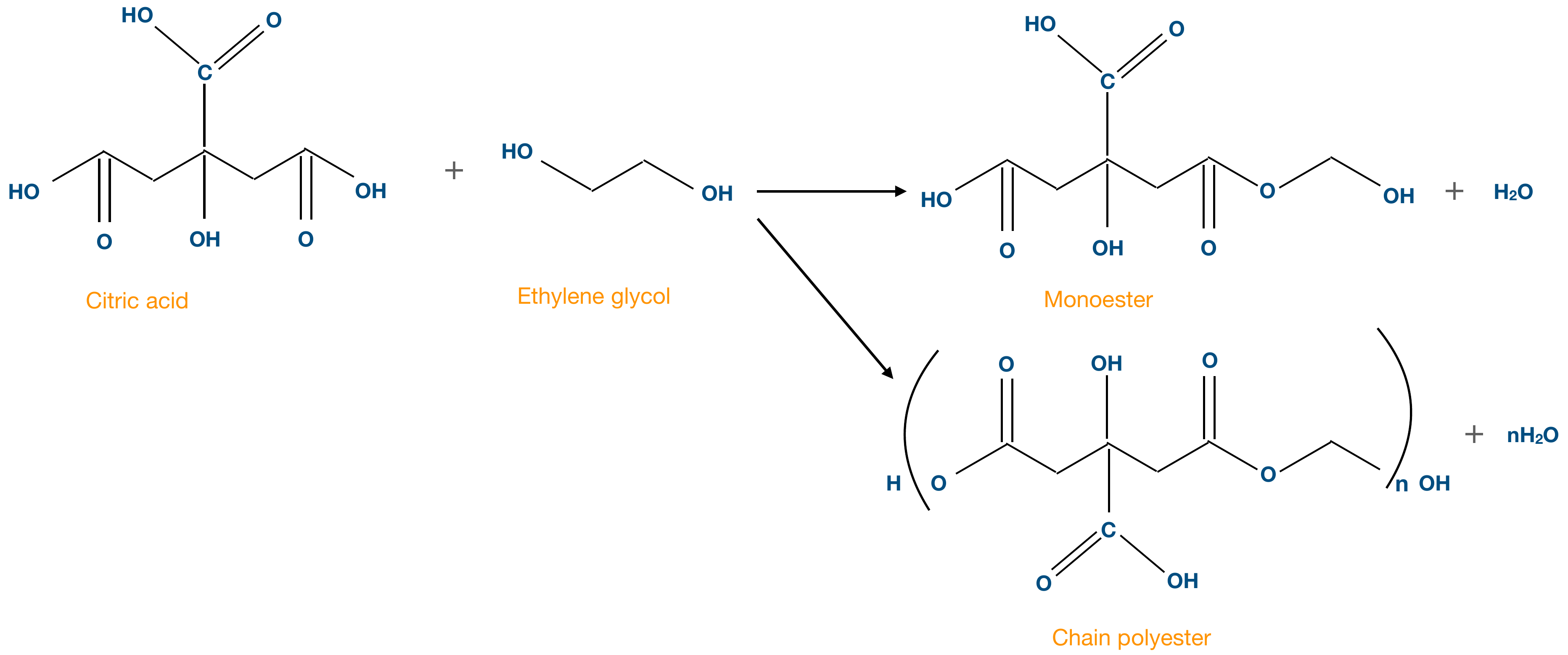}
\caption{Illustrating the polyesterization procedure  at 85$^\circ$C of the citric acid with ethylene glycol to establish chelating agent (including dehydration process) while ethylene glycol enhances the viscosity of the precursor.  
\label{fig:ester}}
\end{figure*}

\subsection{Atomic-level mixing sol-gel  synthesis approach}

In general, 
the chemistry-driven sol-gel reaction represents a notable synthesis technique in producing a wide range of glassy and ceramic materials \cite{sol-gel-found-1996}. 
Although the ordinary procedure  consists of multiple chemical transformations from a liquid solution to a gel state with  
a subsequent post-heat treatment, 
the Pechini route involves mainly distributing metallic cations with respect to chelating polymeric agents such as citric acid, lactic acid,   ethylenediaminetetraacetic acid, etc. 
Variables within the  polyesterification process are to be controlled, including (i) the chemical behavior of metallic chelation,  
(ii) the  degrading level of free chelating compound, e.g., citric acid, 
(iii) the polymeric crosslinking stage, which may affect the uniformity of the cation distribution in 
the final gel, 
and 
(iv) the coordination structure of metal-complexes or polymer-metal-complexes in solution  \cite{sol-gel-found-1996, RAJENDRAN1994239}. 
Utilizing the polymeric precursor  Pechini path, we performed in Li-doped Bi(Pb)-2223 ceramic superconductors with a nonstoichiometric formula \ce{Bi_{1.4}Pb_{0.6}Sr_2Ca_2(Cu_{1−x}Li_x)_3O_{10+\delta}}, with $x$ = 0.0--0.20 with 
the 0.05 increment step, and $\delta$ denotes the oxygen surplus content,
 in a twofold  mode. 
 For convenience, we labeled the five samples as Li0, Li5, Li10, Li15, and Li20, corresponding to the respective doping concentrations.

\begin{figure*}[ht]
\centering
\includegraphics[scale=0.23]{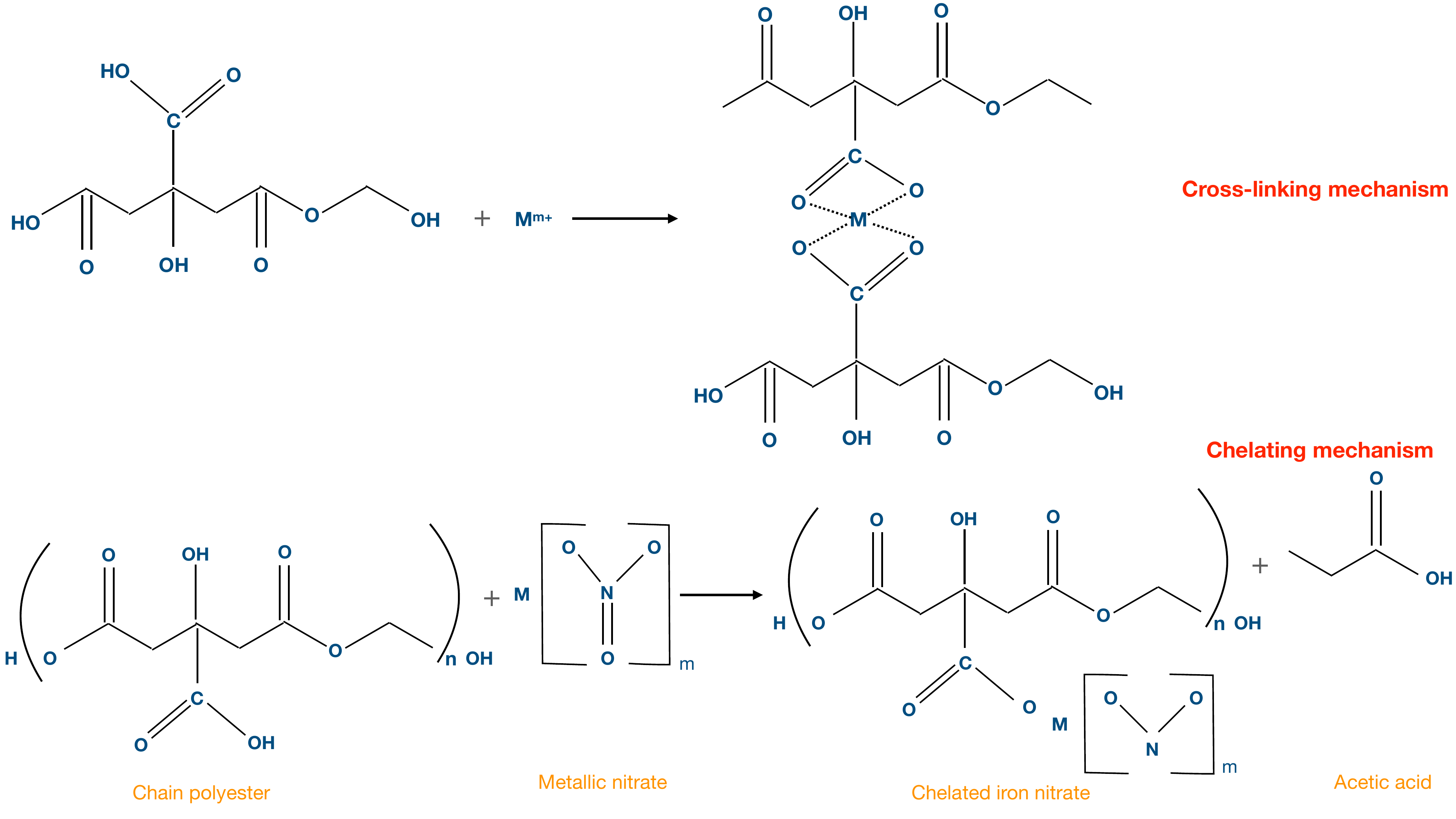}
\caption{  Entrapping cations within polyesterisation by cross-linking and chelating  mechanisms within Pechini solgel route. 
\label{fig:Pechini}}
\end{figure*}

\paragraph{Stage 1: polymeric complexity Pechini trajectory}

We employed the elementary recipes from aqueous solutions of high-purity nitrate salts including multivalent Bi$^{3+}$, Pb$^{2+}$, Sr$^{2+}$, Ca$^{2+}$, and Cu$^{2+}$, as well as monovalent doping Li$^+$ cations, whose molar components are equivalent to the stoichiometric formula. 
In preparation, precise molar amounts of  nitrate salts  were dissolved in the deionized aqueous solution to store  separately, 
and the subsequent  solution was mixed together to obtain a light-blue precursor. 
Additionally, the organic polymer precursors, so-called $\alpha$-hydroxycarboxylic multicaborxyl-hydroxyl citric acid (CA), represent backbone chelating ligands for metallic 
cations, while polyhydroxyl  ethylene glycol serves as a crosslinking agent to
establish a polymeric resin in the molecular regime 
or chain of polymer in the esterification process, as a result, the viscosity of precursor solution  constantly increases  \cite{PSG-1, sol-gel-found-1996, KAKIHANA1998129} (illustrated in Fig.~\ref{fig:ester}).
Upon we mixed the  metallic nitrate salt solution with polymer precursor following a ratio CA : EG  = 1.5 : 1, ratio CA : EG : M$^{n+}$ = 1.5
: 1 : 1 (M$^{n+}$ = Bi$^{3+}$, Pb$^{2+}$, Sr$^{2+}$, Ca$^{2+}$, Cu$^{2+}$)  and stirred at 85$^\circ$C on  the magnetic stirrer  hotplate. 
The water  evaporated from
the mixed solution at 85$^\circ$C, 
and this favors the organic chemical reaction to the right in the formation range from monomer to polymer (Fig.~\ref{fig:ester}). 
However, with a low temperature of 85$^\circ$C, the solution prefers to form lower degree of polymeric resin \cite{sol-gel-found-1996}. 
%

The solution was intensively conglomerated and agitated at 90$^\circ$C,
and it was gradually added to the
precursor solution to prepare stable metallic chelate complexes. 
Immobilization of metal complexes in such rigid organic polymer networks reduces
segregation of particular metal ions, ensuring compositional homogeneity \cite{PSG-1, sol-gel-found-1996, KAKIHANA1998129}.
In further heating to 120$^\circ$C for 5 hours, the viscous solution was converted into a dark blue foam-like block on the bottom of the cup, accompanied by a transparent gel without any precipitation  
as well as completely eliminate  water and organic impurities. 
Metallic ions 
therefore  play a  role as cross-linking agents 
between polymers or as  a part of a chelating nitrate complex polymer as illustrated in Fig.~\ref{fig:Pechini}.  
The effect of adding a monovalent Li$^+$ cation to the system will be tested to determine which favorable mechanism drives the Pechini formation of cross-linking or chelating pathways. 
Since Li$^+$ cations are involved, cross-linking will be more preferred under our fabrication conditions. 

%

\paragraph{Stage 2: Gel pyrolysis and high-$T_c$ phase growth}
The solid precursors were pyrolyzed at
650$^\circ$C for 10 hours to establish ultrafine powder precursors of the compound, in which  
water and other organic chelating solvents are entirely burned out in this step. 
After this stage, 
the calcined powders were ground and pressed into pellet bars under hydrostatic pressure of five  tons/cm$^2$. 
The subsequent samples were placed inside a quartz tube furnace and sintered under atmospheric conditions at a temperature of  850$^\circ$C for seven days. 
Compared to the standard solid-state reaction method, this stage is quite analogous  in fabricating the standard high-$T_c$ BSCCO superconductors \cite{PRM-2024-Do, RSI-2025-Do, Man-JKorean}.

\subsection{Sample Characterization }

In order to manifest the formation of the high-$T_c$ Bi-2223 phase prepared by the Pechini sol-gel synthesis, 
we performed X-ray diffraction (XRD) spectroscopy with CuK$_\alpha$  radiation to observe the crystallization of the samples as well as  high-$T_c$ phase formation and 
the effect of the Li doping content on the powder form. 
Furthermore, we characterize the surface morphology, grain size, and an insightful crystalline growth to capture for the samples utilizing Jeol-5410-LV Scanning Electron Microscope (SEM) over different scales.

Continuing, we carved  specimens into square-bar shapes with the approximate  dimensions of $2\times 2 \times 12\text{mm}^3$ mimicking a wire.
Subsequently, we carried out the measurements by attaching Li-doped Bi-2223 specimens on the cold finger of a Helium closed-cycle system (CTI Cryogenic 8200) and a lock-in amplifier in the range of 20--300 K. 
We conducted the temperature dependence of  dc transport $\rho (T) $ and complex ac susceptibility $\chi (T)$ measurements ($\chi = \chi^{\prime} + i \chi^{\prime \prime}$ with in-phase $\chi^{\prime}$ and  out-of-phase $\chi^{\prime \prime}$ components) by an advanced four-probe and lock-in amplifier techniques, respectively.
For supplemental details, we employed phase-sensitive detection within a dual-phase log-in amplifier, which allows us to measure ac susceptibility correctly and simultaneously both  the imaginary and real parts  at each temperature. 
Whereas both coils are  directly wrapped around  the specimen which minimizes temperature and phase deviation, 
the primary coil was fed by 
a sinusoidal ac current with controlled magnitude and frequency to generate a homogeneous ac magnetic field and an induced voltage was measured  on the  secondary coil.  

\subsection{Theoretical analysis of ac susceptibility}

It has been stated that complex ac susceptibility being a powerful tool for  
studying various superconductors and  magnetic materials, 
where the frequency dependence of $\chi^{\prime}$ and $\chi^{\prime\prime}$ gives important 
information on the magnetic structure and the mechanism 
of magnetization processes \cite{Topping_2019}. 
The major frontier of high-temperature cuprate  superconductors consists of anisotropic grains that are separated by grain boundaries that act as weak links of the Josephson network \cite{Muller, Muller-PRB-1991}. 
Under a harmonic magnetic field $B_{ac} = B_0 \cos(\omega t) $ as a driving force, the response function of  
the complex AC susceptibility components  consist of real and imaginary complementary parts:
\begin{align}
    \chi^\prime & = \frac{1}{\pi B_a} \int_0^{2\pi} M(\omega t) \cos(\omega t) d(\omega t),  
    \\
    \chi^{\prime\prime} & = \frac{1}{\pi B_a} \int_0^{2\pi} M(\omega t) \sin(\omega t) d(\omega t).  
\end{align}
The response of such a granular superconductor to a weak magnetic field is determined by the diamagnetic 
response of the superconducting grains and the pinning  Josephson vortices which form along grain boundaries, producing an intergranular field \cite{RMP-2002-Hilgenkamp}. 
The effect of intergranular flux creep is incorporated into the critical state model and explains 
the observed frequency dependence of the ac susceptibility. 
%

%
%
%
%
%

Flux pinning represents the unique characteristics of type II  hard superconductors \cite{Anderson-PRL-1962} 
in which the defects or grain boundary impurities entrap the magnetic flux, especially in sintered cuprate superconductors.
That opens a possibility  utilizing  the effect of frequency to elucidate the properties of flux trapping in high-$T_c$ superconductors. 
%
In the case of a long cylindrical sample with radial coordinate $r$, 
where the applied field is parallel to the cylinder axis, the critical state equation for the intergranular magnetic field profile 
\begin{equation}
    \frac{dH(r)}{dr} = \pm \frac{F_p}{\lvert B(r) \rvert},
\end{equation}
where $F_p$ is the intergranular pinning force density of Josephson vortices.
Using the Anderson flux creep model, the pinning force density is expressed as
\begin{equation}
    F_p = \frac{k_BT}{V_b r_p} \arcsin{\bigg [ \frac{v_h}{v_0} e^\frac{U(T,H)}{k_bT} \bigg ]},
\end{equation}
where $U(T,H)$ names the average intergranular pinning potential, $r_p$ typifies the potential range and $V_p$ is the flux bundle volume (Josephson vortices along grain boundaries), which is assumed to contain a single flux quantum $\Phi_0$, $v_0$ the attempt frequency for hopping and $v_h$ is the ``minimal-observable'' flux line hopping rate.
The minimal observable hopping rate, $v_h$, for an applied steady magnetic field, is the inverse of is behaved 
\begin{equation}
    v_h \approx \frac{8 \Phi_0 H_{ac} \bar{R}_g^2}{Rk_BT} f.
\end{equation}
In the case of a long cylindrical sample with radius $R$ and an applied magnetic ac field of amplitude $H_{ac}$, Boltzmann  constant $k_B$  and frequency $f$, which the activation energy will be established.

\begin{figure}[ht!]
\centering
\includegraphics[scale=0.28]{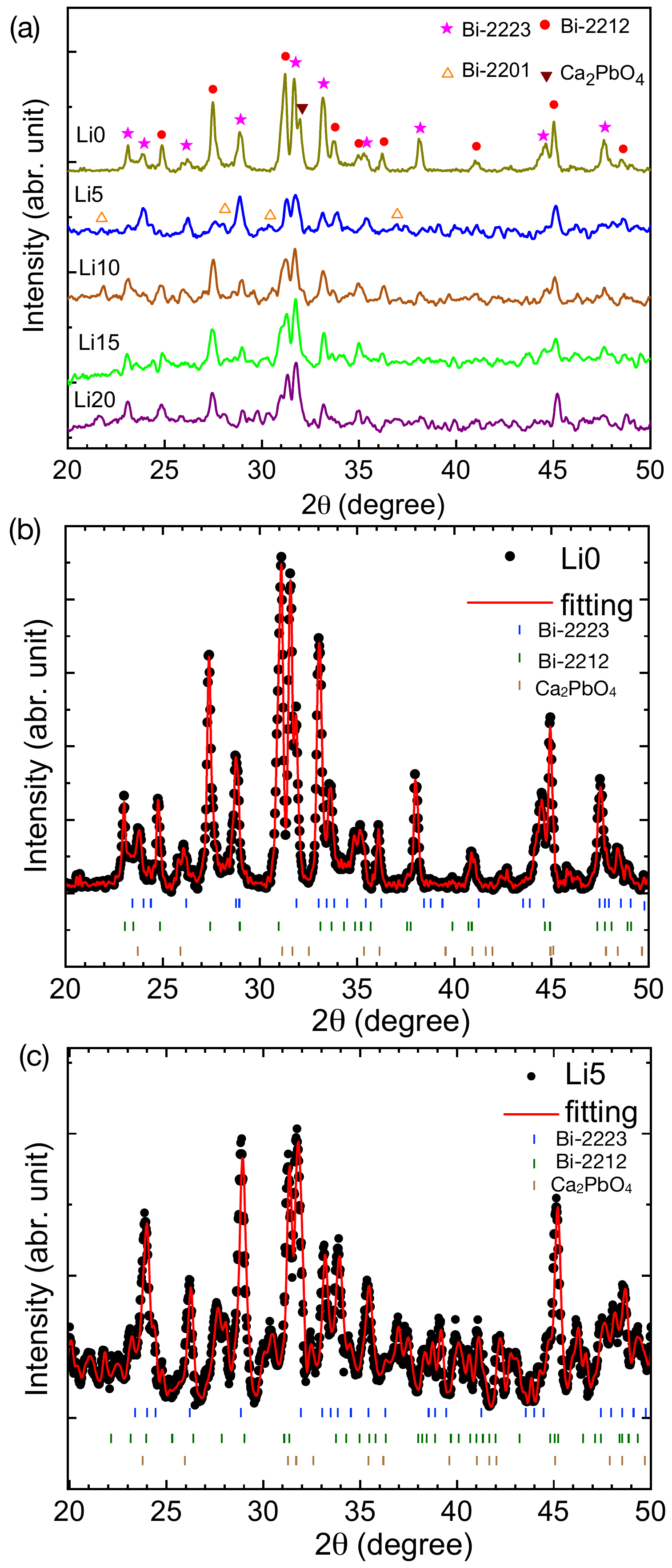}
\caption{ (a) Series of X-ray diffraction patterns of    Bi$_{1.4}$Pb$_{0.6}$Sr$_2$Ca$_2$(Cu$_{1-x}$Li$_x$)$_3$O$_{10 + \delta}$   powders at room temperature 
with respect to Li-doping  content up to 20 at.~\% replacement.
Apparently, the samples consist of  high-$T_c$ triple-cuprate phase as well as amount lower-$T_c$ double-layered Bi-2212 on surrounding boundary of  grains  and other impurities \cite{TAMPIERI2000119}. 
Rietveld refinement for (b) Li0 and (c) Li5 samples based on MAUD software. 
\label{fig:XRD}}
\end{figure}

\section{Results and Discussions}
\label{app1}

\begin{table*}[h!]
 \caption{ Summary of various sol-gel approaches to synthesize Pb-doped Bi-2223 superconductors and their doped counterparts, accompanied by detailed starting precursors, methods, lattice parameter, and    $T_c$ values. (The abbreviation terms are referred as CA: citric acid, EDTA: ethylenediamine tetraacetic
acid, AA: acrylic acid, TF: thin film, SG: sol-gel, and SC: single crystal. 
Some detailed stoichiometric formulas are also noted.) %
   \label{table0}} 
\begin{tabular}{l  l l l l l l l } 
 $x$ & $a$  & $c$  & $T_c$   & $H_\textrm{ac}$, freq.  & Precursors & sintering & Refs.~\\ 
  & \AA & \AA & K & A/m\footnotemark[1], Hz &  & $T_\textrm{s}$($^\circ$C), $t_\textrm{s}$ (h)  &  \\
 \hline
 Li0\footnotemark[2]  &  3.825(6) & 37.147(2) & 107.4  & 12, 10$^4$  & nitrates, CA & air, 850, 168  & Ours 
\\
 Li5 & 3.826(3) & 37.048(6) & 111.4  & 12, 10$^4$  & nitrates, CA & air, 850, 168 & Ours 
\\
 Li10  & 3.831(1) & 37.113(2) & 107.6 & 12, 10$^4$ & nitrates, CA & air, 850, 168 & Ours  
\\
Li15 &  3.810 & 37.139(0) & 83.5  & 12, 10$^4$  & nitrates, CA &  air, 850, 168 & Ours 
\\
Li20 & 3.835(5) & 37.099(5) & N/A & --  & nitrates, CA  & air, 850, 168 & Ours
\\
Bi-2223 & 3.835 & 37.074(4) & 110.9 & 64, 333  & nitrates, EDTA & air, 850, 100 &\cite{FALLAHARANI2022163201}
\\
Bi-2223+TiO$_2$(0.2\%) & 3.833 & 37.069(1) & 106.4 & 64, 333  & nitrates, EDTA & air, 850, 100 &\cite{FALLAHARANI2022163201}
\\
Bi-2223+TiO$_2$(0.4\%) & 3.832 & 37.064(2) & 103.8 & 64, 333  & nitrates, EDTA & air, 850, 100 & \cite{FALLAHARANI2022163201}
\\
Bi-2223\footnotemark[3] & -- & -- & 108.2 & 200, 333  &  carbonates, EDTA & air, 835, 170 & \cite{Shamsodini_2019}
\\
Bi-2223 & -- & -- & 107.1 & 200, 333  &  nitrates, EDTA & air, 835, 170 & \cite{Shamsodini_2019}
\\
Bi-2223/CdO (2\%) & -- & --- & 109 & 5, 333 & acetates & air, 845, 64 & \cite{Cd-doped-Bi2223}
\\ 
Bi2223 (2\%) & -- & --- & 108 & 8, 1 & acetates, EDTA & air, 830/840 & \cite{RUBESOVA2012448}
\\ 
Bi-2223\footnotemark[4] (TF) & -- & 37.1 & 106 & --  &   acetates,  AA & air, 860, 170 & \cite{Shen_2024}
\\
Bi-2223 (SG+SC) & 3.859 & 37.19(1) & 111 & 160, 0  &   acetates,  AA &  O$_2$, 500, 240 & \cite{Pb-doped-Bi2223}
\\
\hline
\hline
\end{tabular}
\footnotesize{$^1$: With measurements using a magnetic field by Oe unit: 79.6 A/m = 1 Oe} \\ 
\footnotesize{$^2$: Bi$_{1.4}$Pb$_{0.6}$Sr$_{2}$Ca$_{2}$Cu$_{3}$O$_{10 + \delta}$} \\
\footnotesize{$^3$: Bi$_{1.64}$Pb$_{0.34}$Sr$_{2}$Ca$_{2}$Cu$_{3}$O$_{10 + \delta}$} \\
\footnotesize{$^4$: Bi$_{1.9}$Pb$_{0.35}$Sr$_{2.}$Ca$_{2}$Cu$_{3}$O$_{10 + \delta}$}
\end{table*}

Once  the samples were obtained, 
we characterized     
their crystal structure and surface morphology as well as 
probed the superconducting properties of Li-doped Bi-2223 by  dc resistivity and   ac susceptibility at low field equipped by a high frequency regime. 
\subsection{Crystalline and surface morphology}

 Figs.~\ref{fig:XRD}(a)--(c) present  a series of XRD for Li-doped Bi-2223 compounds demonstrating  the major phases of Bi-2223 and 2212 together with the existing residual impurities of the liquid Ca$_2$PbO$_4$ and  minor Bi-2101, together with the Rietveld refinement analysis. 
 That is in analogy to  the existence of  three Bi-based superconducting phases observed in several prior sol-gel  studies \cite{FALLAHARANI2022163201, Cd-doped-Bi2223, Shen_2024, Shamsodini_2019}, 
 but different from traditional solid-state reaction or single crystal growth approaches such that only two major phases Bi-2223 and 2212 \cite{ManP-Ag, PRM-2024-Do, Pb-doped-Bi2223}.
 Nevertheless, the main   diffraction peaks belong to the Bi-2223 phase (Fig.~\ref{fig:XRD}).

For more details,  we typically observe  sharp peaks (2$\theta$ $\sim$ 27.5$^\circ$ and 31.2$^\circ$) of the double-layered Bi-2212 phase (58.5~\% contribution) with only 36.7~\% of Bi-2223 in the case of the undoped Li0   sample [Fig.~\ref{fig:XRD}(a)-(b)]. 
A remark is that the single sintering step at low temperature 850$^\circ$C (below the eutectic point of the Bi-2223 superconductor at 870$^\circ$C \cite{PMajewski_1997}) prepared even by an atomic mixing sol-gel route is   not considered appropriate   to grow pristine Bi-2223, 
and  this has also been proven by recent fabrication of a higher sintering temperature $T_s$ = 860$^\circ$C, providing the highest quality thin film  \cite{Shen_2024}. 
However, with the smallest doping of Li with 5 at.~\% into the framework, this sintering temperature becomes suitable to grow the Bi-2223 phase (62.9~\% contribution) with  sharp peaks at 2$\theta$ $\sim$ 28.5$^\circ$ and 32$^\circ$. 
That reaffirms the reduction in the sintering temperature in Li-doped Bi-2223 samples, which complies with previous triple-layered superconductors by prior whisker fabrication and conventional solid-state synthesis \cite{Matsubara, MatsubaraI, PRM-2024-Do}. 
In heavier Li-doping contents, the diffraction peak (typical 2$\theta$ $\sim$ 28.5$^\circ$) of Bi-2223   fade rapidly, that may regard  the difficulty in synthesizing monovalent Li$^+$  doping by the Pechini so-gel route and
as such will be elucidated in further characterizations.

\begin{figure*}
\centering
\includegraphics[scale=0.24]{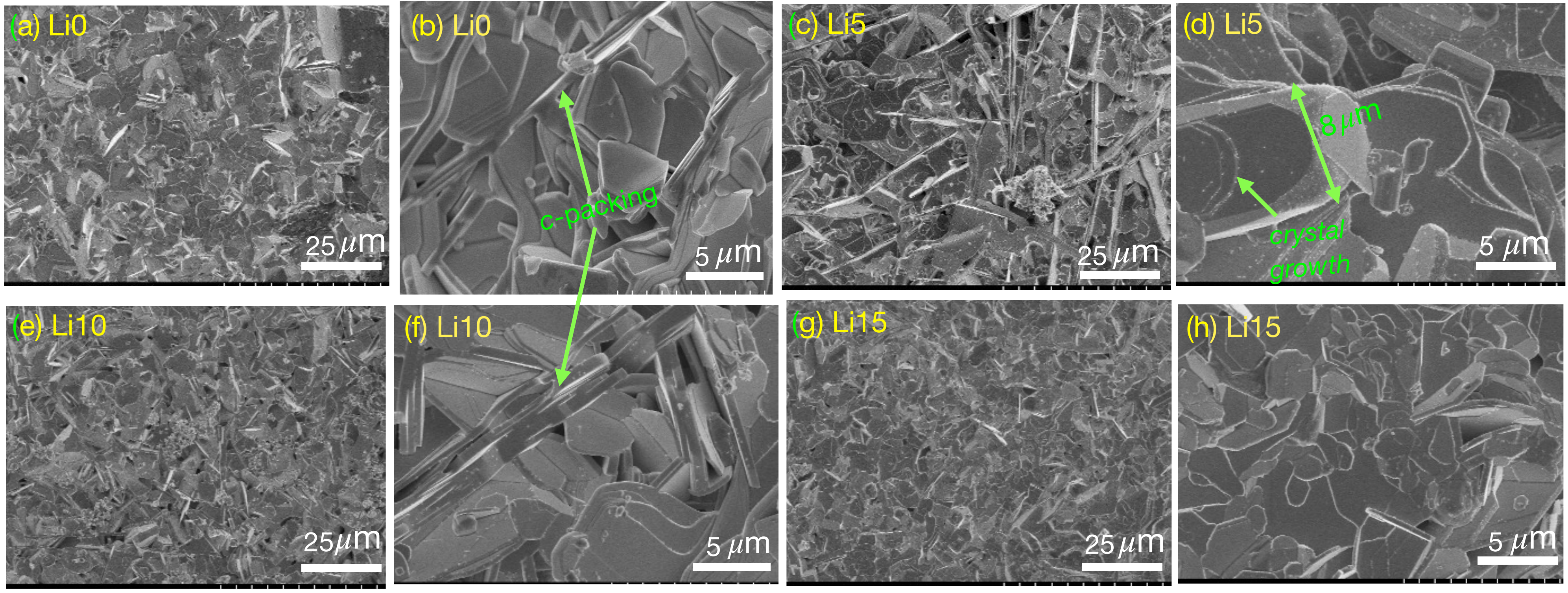}
\caption{ Surface topography of     Bi$_{1.4}$Pb$_{0.6}$Sr$_2$Ca$_2$(Cu$_{1-x}$Li$_x$)$_3$O$_{10 + \delta}$   ceramic superconductors  
in terms of Li-doping  content with (a)--(b) for Li0 sample, (c)--(d) for Li5, (e)--(f) for Li10 and (g)--(h) for Li15 sample. 
%
\label{fig:SEM}}
\end{figure*}

As in the investigation, Bragg peaks attributed to the Li$_2$O crystal phase have not appeared  in the XRD spectra even with high concentration of Li$_2$O. 
It may be noticed that the high  content of Li$_2$O altered the 
homogeneity of the ephemeral liquid forming, the reaction rate, and 
as a result, the formation rate of the Bi-2223 phase.
No observation of peaks related to Li-containing compounds, which means that Li$_2$O intercalated to the superconducting phases, either low or high, even with a large amount of impurities up to 20 at.~\% away from Cd-substituted for Ca in Bi-2223 compounds  prepared by solgel synthesis \cite{Cd-doped-Bi2223}.

 With the spotlight on the Bi-2223 phase, 
 based on the pseudo-tetragonal  structure I4/mmm group of the Bi-2223 phase, 
the lattice parameters $a$ = 3.825--3.835 \AA~and $c$ = 37.048--37.147 \AA~are well-matched to known values ($a$ = 3.832--3.859 \AA~and $c$ = 37.064--37.19 \AA~\cite{FALLAHARANI2022163201, Pb-doped-Bi2223}),     
as we present in Table~\ref{table0}. 
This is due to the fact that the atomic radius of Li$^+$ (0.73--0.76 \AA)  is quite equivalent to the Cu$^{2+}$  (0.71--0.73 \AA) cation. 
Overall, the parameters are very analogous to the existence of three Bi-based superconductive  phases observed in other studies (Table~\ref{table0}) \cite{FALLAHARANI2022163201, Shen_2024}.

We now consider the surface morphology with a clear grain size in large and small measuring  scales  
(Fig.~\ref{fig:SEM}). 
At a large scale of 25 $\mu$m, we view a wide pervasive surface topography of samples accompanied by dense plates and needle shapes upon exposing heat and oxygen during the long-time   sintering process. 
They behave     somewhat more porous compared to the previous Li-doped in Bi-2223  prepared by a traditional solid state reaction method \cite{PRM-2024-Do} [Figs.~\ref{fig:SEM}(a), (c), (e) and (g)].
However, 
grains prepared from sol-gel are  distinct with a sharp grain boundary,  
while it smooths and solidifies from solid reaction counterpart at the equivalent sintering condition of 850$^\circ$C \cite{PRM-2024-Do}.
A granular structure with flaky plate-like layers oriented in facet random directions is observed in the  samples. 
In particular, Li0 and Li10 pronounce a very similar surface morphology accompanied by a random plate crystal [Figs.~\ref{fig:SEM}(a) and (e)], particularly  
showing thin $c$-packing granular plates with a thickness of 1--2 $\mu$m normal to the surface [Figs.~\ref{fig:SEM}(b) and (f)], and that indicates the analogous superconducting properties in the next section. Notably, 
we capture the obvious crystalline growth on the sample surface while exposing heat and oxygen ions during the sintering process which 
show the preferential growth of the CuO$_2$ plane along the $ab$-plane [Fig.~\ref{fig:SEM}(d)]. 
Their surface
morphology indicates the strong dark-gray color of micro-size
crystal plates (3–5 $\mu$m) and  polygonal disks  (5–10 $\mu$m).
Compared to the previous works, sintering at lower temperature 835$^\circ$C, grain size of particles is smaller \cite{Shamsodini_2019}, or thin film under dip coating, the grain boundary is elusive \cite{Shen_2024}. 
Doping Cd for the Bi-2223 system even creates multiple pores on the grain as a consequence of degrading superconducting properties \cite{Cd-doped-Bi2223}. 
%

\subsection{The effect of Li$^+$ substitution on superconducting properties by DC transport and ac susceptibility}

\begin{figure}
\centering
\includegraphics[scale=0.26]{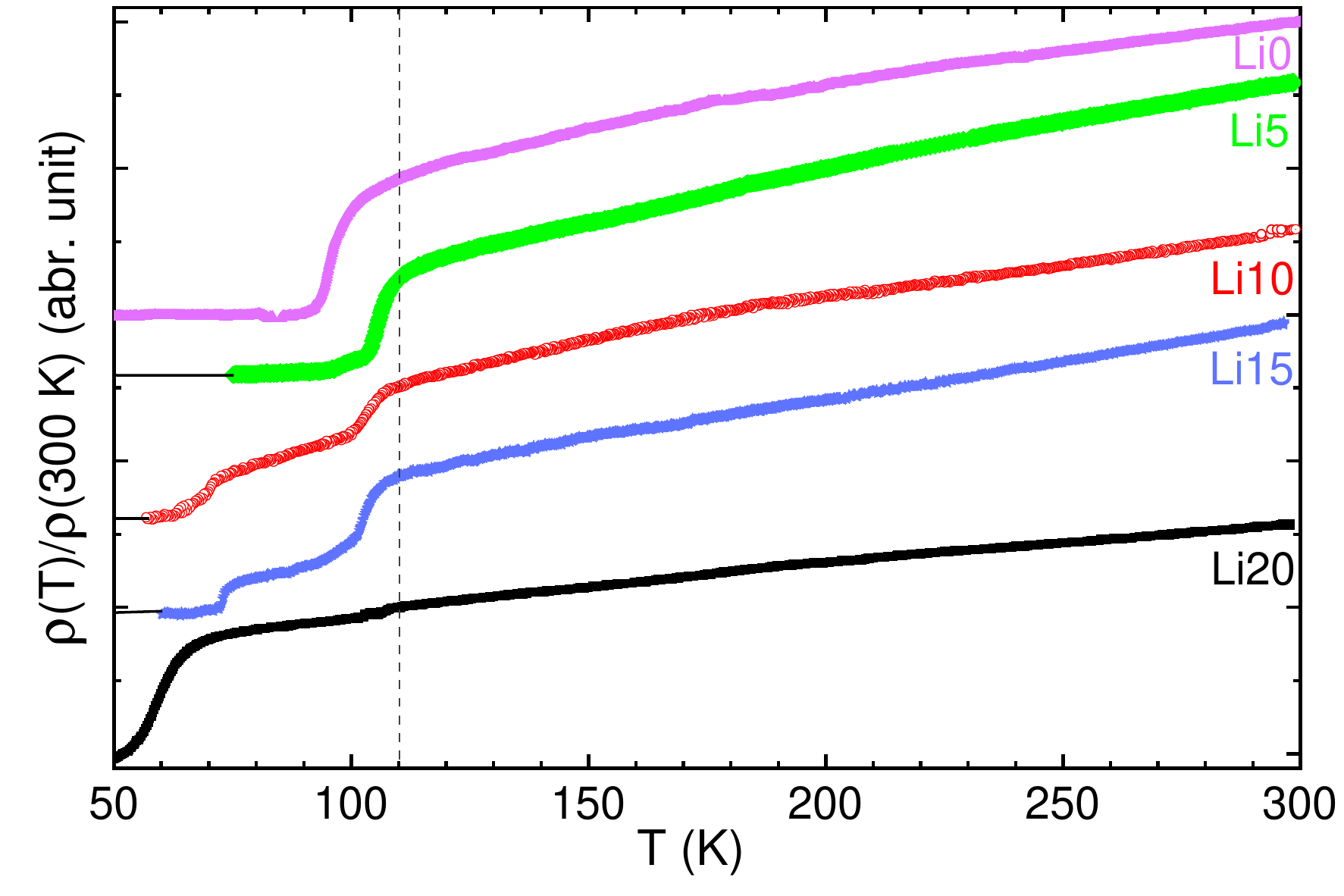}
\caption{ Entire resistivity versus temperature $\rho(T)$ of  Bi$_{1.4}$Pb$_{0.6}$Sr$_2$Ca$_2$(Cu$_{1-x}$Li$_x$)$_3$O$_{10 + \delta}$   (with $x$ = 0.0--0.20) compounds measuring in the temperature range of 50--300 K.
\label{fig:RT}}
\end{figure}

DC transport measurements display  normalized resistivity in the range of 50--300 K in which the shape varies following  the Li-doping content in Fig.~\ref{fig:RT}.
Above the $T_c = 110$ K of Bi-2223, there is no observation of an abnormal point with linear resistance reduction versus temperature  (see Fig.~\ref{fig:RT}), providing no evidence of  charge density wave phases or impurities,   which belongs to strange metallic behavior, e.g., transition from metal to insulators in heavy Eu doping \cite{YILDIRIM2013526}. 
For samples $x$ = 0--0.15, the resistivity shows a transition of 110 K with respect to the triple-layered Bi-2223 phase, but the sharp transition is only for small doping concentrations (up to 5 at.~\% Li-substitution).  
The heavy replacement of 10--20 at.~\% demonstrates multiple transitions, including the amount of low-$T_c$ single- or double-layered phases (agreeing with the above XRD study) that breaks the Bi-2223 phase formation in analogous to  
  traditional solid-state reaction preparation \cite{PRM-2024-Do}, 
 synthesizing Li-doped Bi-2223 by the Pechini sol-gel route faces  some difficulties and challenges in establishing the high-$T_c$ phase. 
 The DC transport manifests that  Li5  sample gives the best for resistivity shape  measurement with zero resistance temperature of 100 K. 
 That is complementary  to controlling the  sintering temperature from 845--860$^\circ$C for thin film fabrication in which $T_s$ = 860 gives the optimal  result \cite{Shen_2024},  while intercalating TiO$_2$ into Bi-2223 continuously degrades the  interlinking particles as the TiO$_2$ content increases \cite{FALLAHARANI2022163201}.

\begin{figure}[htb!]
\centering
\includegraphics[scale=0.26]{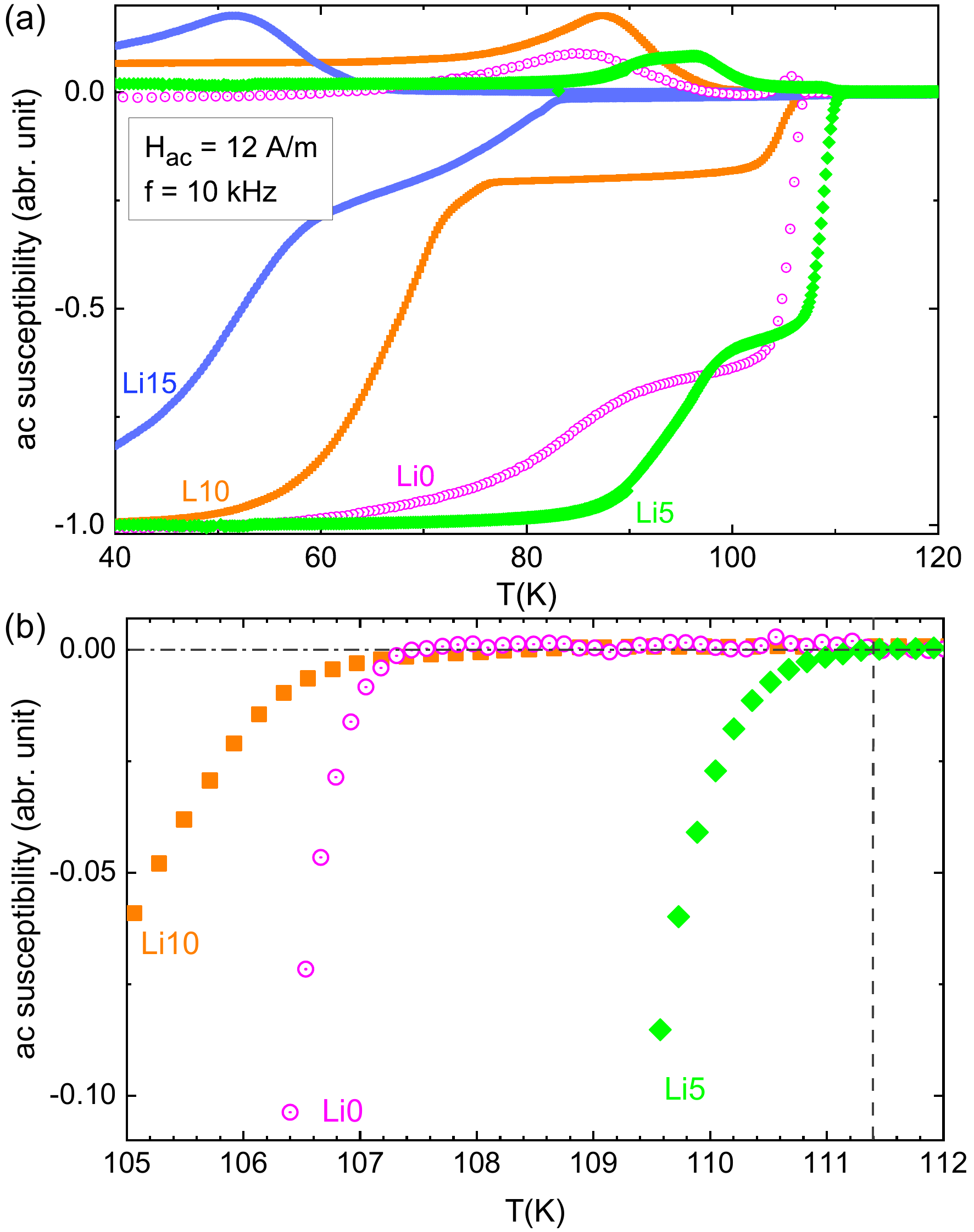}
\caption{ (a) The magnetic susceptibility of Li-doped Bi$_{1.4}$Pb$_{0.6}$Sr$_2$Ca$_2$Cu$_3$O$_{10 + \delta}$ ($x$ = 0--0.15) was synthesized by the Pechini sol-gel synthesis and measured at the field amplitude  $H_\textrm{ac}$ = 12 A/m  and frequency $f$ = 10 kHz. (b) Zoom in the transition region of real susceptibility around 110 K.   
\label{fig:2}}
\end{figure}

In addition to DC transport, we further characterize high-$T_c$ materials by ac susceptibility 
which makes it a more versatile approach  \cite{AC-Suscept, CELEBI1998131, AC-Topical-review} 
to probe the detailed  superconducting transition as well as  the intrinsic microstructure.
In Fig.~\ref{fig:2}, we perform ac susceptibility measurements  for  both samples through utilizing a small applied  field of $H_\textrm{ac}$ = 12 A/m and a frequency of 10$^4$ Hz for Li-doped samples  
in the range of 40--120 K, and each real susceptibility includes a two-step transition process of intragrain and intergrain drops.
The samples $x$ = 0 and $x$ = 0.05  behave quite similarly to intragrain and intergrain transitions, but a higher Li-doping content deviates from them.
In the circumstance of $\chi^{\prime} = -1$ (Li0, Li5 and Li10 samples in Figure~\ref{fig:2})  and simultaneously, the imaginary component $\chi^{\prime \prime}$ approaches 0,  superconducting grains connect  with each other via an intergrain Josephson junctions, forming a whole superconducting block in which  the external magnetic field is completely shielded by the circulating persistent supercurrent around it, known as the Meissner effect \citep{Muller, AC-Suscept, AC-Topical-review}.
%
Therefore, 
this characterization is consistent with the DC transport measurement presented above. 

\begin{figure*}[htb!]
\centering
\includegraphics[scale=0.3]{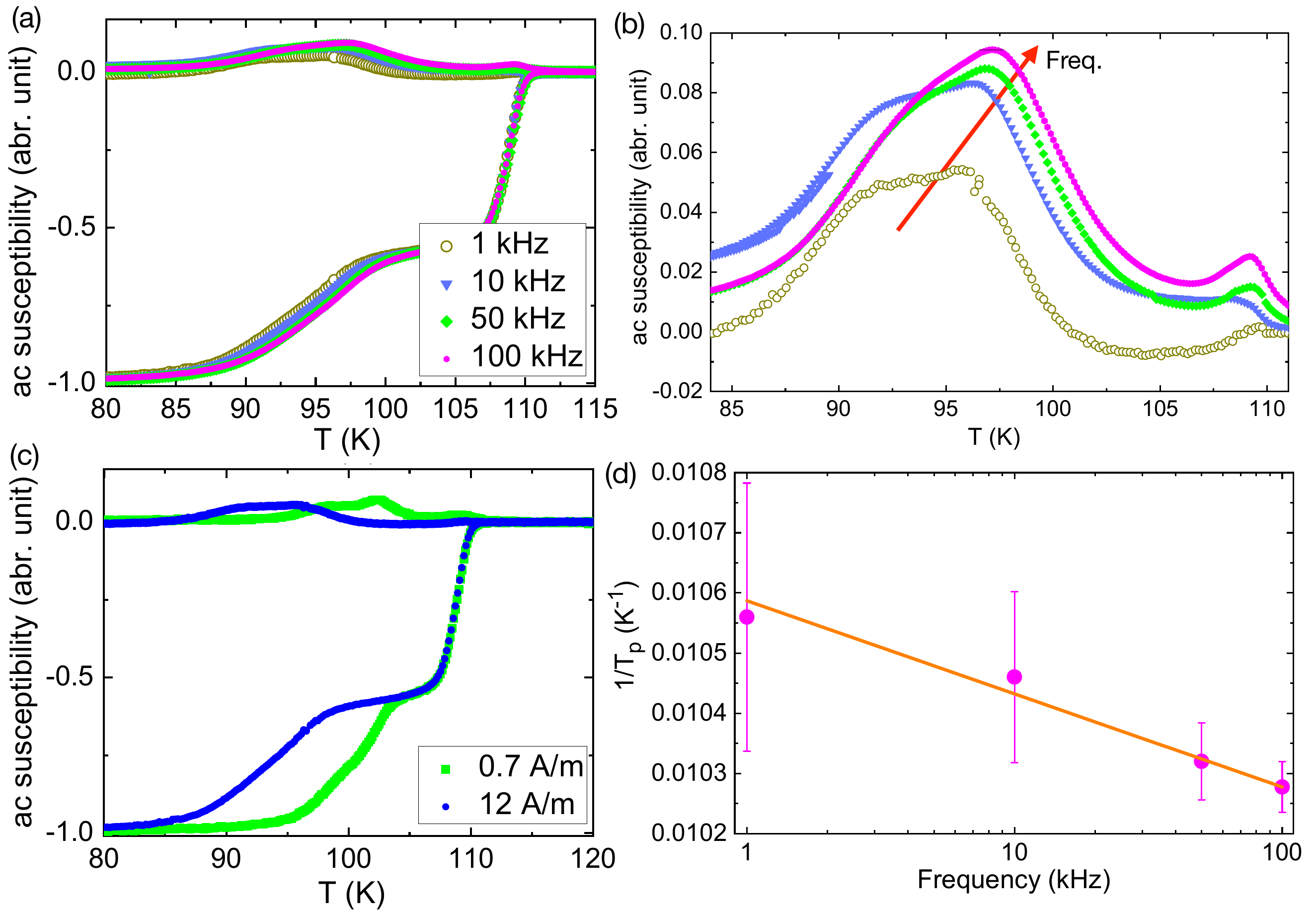}
\caption{ The magnetic susceptibility of Bi$_{1.4}$Pb$_{0.6}$Sr$_2$Ca$_2$Cu$_{2.85}$Li$_{0.15}$ O$_{10 + \delta}$  (a) measured at the field amplitude  $H_\textrm{ac}$ = 12 A/m  and frequency $f$ = 1-100 kHz, (b) zoom in detailed intergrain peak region, (c) measured at the different field amplitudes of 0.7 and 12 A/m and (d) plotting $T_p$ versus frequency.  %
\label{fig:3}}
\end{figure*}

Upon doping Li into the Bi-2223 framework, the critical temperature advances from 107.4 K (Li0) to the peak of 111.4 K (Li5) and down to 107.6 K (Li10).
Once we juxtapose the values   by our Pechini  route with other sol-gel  approaches to elucidate the role of starting precursors and heat treatment procedure on superconducting properties.
Other newly sol-gel fabrications have been applied to synthesize Bi-2223 compounds, such as starting nitrates or carbonates precursor with sintering temperature of 835$^\circ$C giving $T_c$ = 107.1--108.2 K \cite{Shamsodini_2019},  
intercalating  TiO$_2$ with sintering at 850$^\circ$C producing  $T_c$ = 103.8--110.9 K \cite{FALLAHARANI2022163201} and sol-gel following a single crystal growth $T_c$ = 111 \cite{Shamsodini_2019} (see the list on the  Table~\ref{table0}).
We remark that the starting salt precursors have no  strong effect on synthesizing high-$T_c$ phase with starting from nitrates, carbonates or acetates, $T_c$ is about 106--111 K. 
Lower sintering temperature influences the grain connectivity as well as governing lower the critical temperature.  
Alternation of components within the framework has a significant affect on  preparation process and superconducting properties. 
For instance, substituting Li$^+$ for Cu$^{2+}$ within  Bi-2223 materials gives  the critical temperature of our specimens a peak at $T_c = 111.4 \pm 0.2$ K at x = 0.05, 
while intercalating TiO$_2$ into the system, the critical temperature constantly reduces 110.9--103.8 K (Table~\ref{table0})  at quite similar starting precusors and sintering process. 
Thus, our specimen synthesized by the advanced Pechini method underscores slightly higher $T_c$ by 0.5--1.5 K than preceding fabricating methods and sharp transition.   
With the spotlight on  this impression,  
 we pioneer the combination of appropriate doping with atomic mixing and long time sintering that play a primary role in fabricating the high quality of Bi-2223 compounds. 
%


\subsection{Low-field ac susceptibility  approaching  flux creep property at intergranular boundary}

Considering  an ac susceptibility as a function of temperature, frequency and magnetic field, we investigate the Josephson weak-link effect in   Li-doped Bi-2223 superconductors  by tuning   frequency and  ac field.   
%
%
%
In this case, 
we select $x$ = 0.05 sample to perform and characterize the flux creep effect by implementing a wide frequency range of  1--100 kHz at a low ac magnetic field of 12 A/m  [depicted in Fig.~\ref{fig:3}(a)]. 
Below the critical temperature, the imaginary part of susceptibility exhibits a broad peak that is attributed to hysteresis losses at the grain boundaries and  
 a shift of the coupling peak tends to higher temperature with increasing frequency in
the 1-100 kHz range [Fig.~\ref{fig:3}(b)]. 
Although the intrinsic internal grain transition is pronounced intact by both the magnetic field and frequency, 
the intergrain transition has been heavily affected by the field amplitude [Figs.~\ref{fig:3}(a) and (c)].
The shift in the susceptibility curves was interpreted
as arising from flux creep at the grain boundaries, which was first predicted by Anderson \cite{Anderson-PRL-1962}, following the Abrikosov theory, and heavily developed by M\"uller \cite{Muller, Muller-PRB-1991}.  
In that case, the smallest possible breakdown of
superconductivity caused by the motion of a quantum of magnetic flux. 
At appreciately higher temperatures of high-$T_c$  Bi-2223 superconductors, 
thermally activated flux
creep presents a significant effect \cite{PRB-flux-creep1989} and  
the flux motion is resisted by
viscous drag and inhomogeneities in the samples that pin the flux. 
Even with pinning magnetic quantum flux, the resistance is not exactly zero because thermally activated fluctuations can overcome pinning. 

\begin{figure}[htb!]
\centering

\includegraphics[scale=0.22]{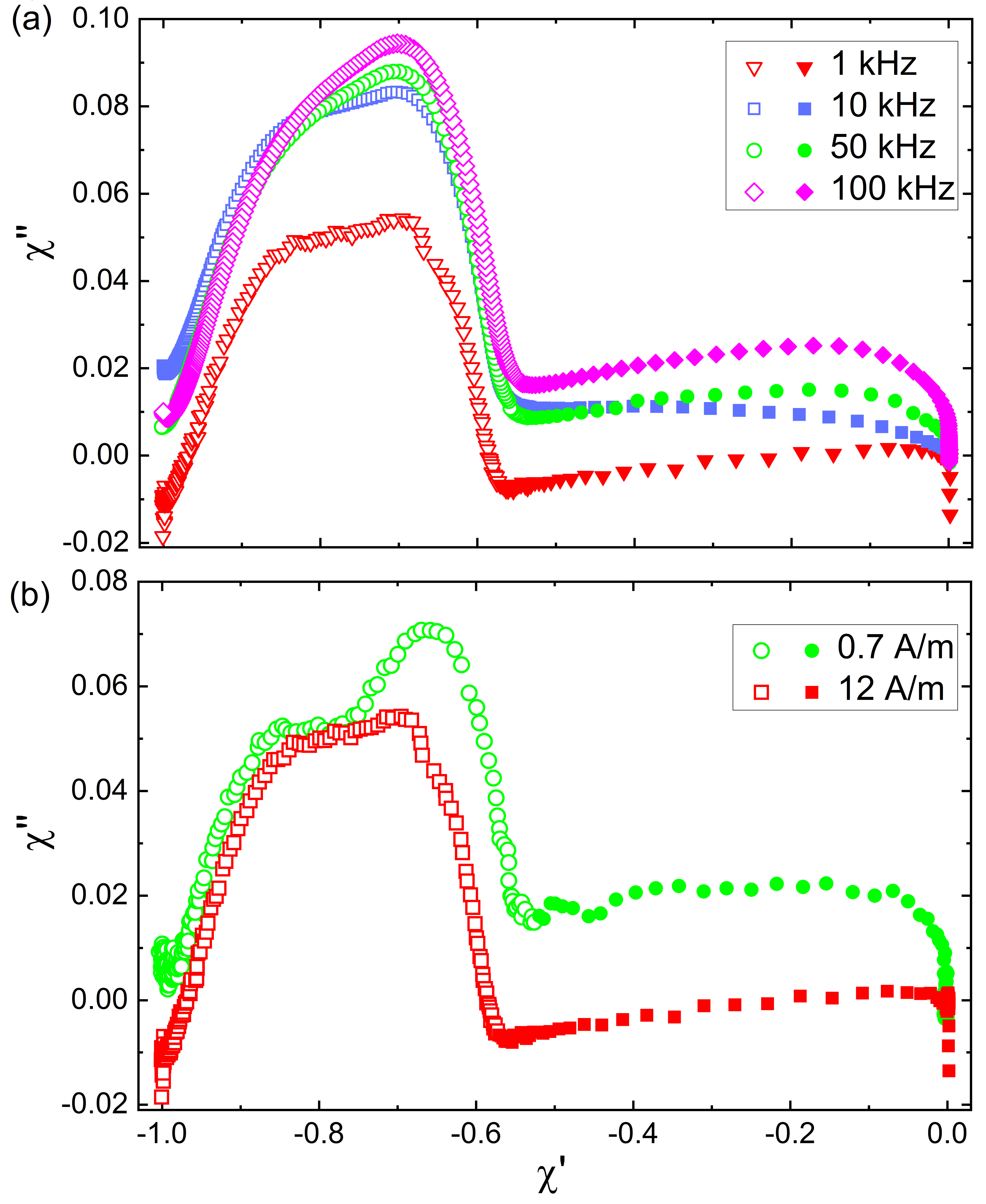}
\caption{ The Cole-Cole plots of Bi$_{1.4}$Pb$_{0.6}$Sr$_2$Ca$_2$Cu$_{2.85}$Li$_{0.15}$O$_{10 + \delta}$ was synthesized by the Pechini sol-gel synthesis and measured at (a) the field amplitude  $H_\textrm{ac}$ = 12 A/m  and frequency $f$ = 1--100 kHz and (b) $H_\textrm{ac}$ =  0.7 and 12 A/m with frequency of 10 kHz.  
\label{fig:4}}
\end{figure}

Anderson suggested that an energy barrier prevented the creep of flux lines \cite{Anderson-PRL-1962, Muller-PRB-1991}. 
The activation energy $\varepsilon_a$ is defined based on the Arrhenius equation as
\begin{equation}
    f/f_0 = \exp(-\varepsilon_a/kT_p), 
\end{equation}
where $f_0$ is a constant. 
In the $\log$-scale [Fig.~\ref{fig:3}(d)], the slope is calculated to be -1.55, then the activation energy is $\varepsilon_a$ = 0.56 $\pm$ 0.06 (eV) at the ac magnetic field of 12 A/m. 
This value is lower than the activation energy provided by 7.0--1.5 eV for the magnetic field range of 0.8--800 A/m \cite{Muller-PRB-1991}.  
In the samples prepared by the Pechini sol-gel synthesis,  smaller energy is consumed to depin the quantum flux in terms of weak-coupling at the grain boundary which will be necessary to further improve.

Finally, the plot of the imaginary part as a function of the real branch within  the harmonic susceptibility   produces the Cole-Cole plot, and this kind of
representation gives additional information complementary to the plot of the separate components as the ac function with respect to the
temperature.
In terms of the Cole-Cole representations [Figs.~\ref{fig:4}(a)--(b)], we exactly demonstrate  two peaks regarding to intragrain (right lower dome) and intergrain (left higher peak) step-like transitions, the frequency on the Cole-Cole plot has much more influences  than separate parts $\chi^\prime$ and $\chi^{\prime\prime}$ as refers to Figs.~\ref{fig:3}(a) and (c). 
The higher frequency makes the plot more apparent   while the higher applied magnitude forms better Cole-Cole plot shape [Figs.~\ref{fig:4}(a)--(b)]. 
 As we observed the peak height of the left dome-shaped curves grows for increasing frequencies, the phenomenon 
is indicated by the formation of a vortex glass phase, specified by a collective flux creep, in contradiction to  the behavior  provided by simulating  the Kim-Anderson flux
creep model \cite{Cole-Cole-PRB, ADESSO2006457}.

\section{Conclusion}

In summary, we provide a concrete interpretation and  synthesis pathway for growing Li-doped Bi-2223 superconductors by Pechini sol-gel methods accompanied by a single-step sintering process. 
Utilizing standard XRD and scanning electron microscopy, we provide insightful crystal growth as well as a slight reduction of  lattice parameters for Bi-2223 compounds upon the  effect of Li-doping.  
Due to the limitations of the  Pechini chelating citric acid in producing samples containing monovalent Li$^+$ substitution, the critical temperature only advanced for small doping levels (0--0.5) with a value of up to 111.4 K, as determined by ac magnetic susceptibility at low fields. 
Compared with other sol-gel Pechini approaches, we have demonstrated a more pronounced atomic mixing synthesis to enhance the superconducting properties of high-$T_c$ Bi-2223. 
The high frequency approach elucidates the activation energy of the M\"uller flux creep model with grain boundary and Cole-Cole plot to interpret double peaks in ac magnetic susceptibility. 

\bibliographystyle{elsarticle-num}
\bibliography{cas-refs}
\biboptions{sort&compress}

\end{document}